\documentstyle[prd,aps,epsfig,floats]{revtex}
\begin{document}
\draft
\wideabs{
\title{HEAT BATH EFFICIENCY WITH METROPOLIS-TYPE UPDATING}

\author{Alexei Bazavov$^{\rm \,a,b}$ and Bernd A. Berg$^{\rm \,a,b}$}

\address{ 
$^{\rm \,a)}$ Department of Physics, Florida State University,
  Tallahassee, FL 32306-4350\\
$^{\rm \,b)}$ School of Computational Science, Florida State 
  University, Tallahassee, FL 32306-4120\\
} 
% (E-mail: berg@csit.fsu.edu)\\ 

% \date{March 7, 2005}
\date{\today }

\maketitle
\begin{abstract}
We illustrate for 4D $SU(2)$ and $U(1)$ lattice gauge theory that 
sampling with a biased Metropolis scheme is essentially equivalent 
to using the heat bath algorithm. Only, the biased Metropolis method 
can also be applied when an efficient heat bath algorithm does not 
exist. For the examples discussed the biased Metropolis algorithm
is also better suited for parallelization than the heat bath
algorithms.
\end{abstract}
% \pacs{PACS: 11.15.Ha, 12.38.Gc; 05.10.Ln, 87.15.-v, 87.14.Ee.}
}
\narrowtext

\section{Introduction}

The possibility of constructing biased Metropolis algorithms (BMAs) is 
known since quite a while \cite{Ha70}. Although they have occasionally 
been used in the statistical physics \cite{Br85} and bio-chemical 
\cite{DeBa96} literature, it appears that practitioners of 
Markov chain Monte Carlo (MC) simulations have not given this topic 
the attention which it deserves. Reasons for this seem to be that 
(a)~general situations for which BMAs are of advantage have not been 
clearly identified and (b)~a lack of straightforward instructions 
about implementing such schemes. 

On the other hand, the heat bath algorithm (HBA) is one of the 
widely used algorithms for MC simulations. It updates a variable 
with the Gibbs-Boltzmann probability defined by its interaction
with the rest of the system (an introduction to HBAs can, e.g., be 
found in Ref.~\cite{bbook}). But, there exist energy functions for 
which an efficient heat bath implementation does not exist. 

In this paper we show that a BMA similar to the one used for the 
rugged Metroplis method of Ref.~\cite{Be03}, can be employed 
whenever one would normally think about constructing a HBA. When 
an efficient heat bath implementation exists, the performance of the 
HBA and the BMA will practically be identical. However, the BMA can 
still be constructed when the inversion of the cumulative distribution,
required for a HBA, is numerically so slow that it is not a suitable 
option. 
% The effort for implementing the BMA exceed only slightly the 
% work needed to program a regular Metropolis algorithm, which is 
% the simplest but often not very efficient scheme.

In the next section we illustrate our general observation for systems 
from lattice gauge theory. 
% because cluster algorithms do not exist for them, so that the 
% heat bath algorithm is indeed the most promising approach. 
Our first 
example is 4D $SU(2)$ lattice gauge theory for which the HBA was 
first introduced by Creutz \cite{Cr80} and improved in 
Ref.~\cite{FH84} and~\cite{KP85}. Our second example is 4D $U(1)$ 
gauge theory.

\section{Biased Metropolis Algorithms and Pure Lattice Gauge Theory}

The action which we consider is
\begin{equation} \label{SUN_action}
  S(\{U\}) = \frac{1}{N_c} \sum_{\Box} {\rm Re}\,
  {\rm Tr} \left( U_{\Box}\right)\,,
\end{equation}
$ U_{\Box} = U_{i_1j_1} U_{j_1i_2} U_{i_2j_2} U_{j_2i_1}$, where 
the sum is over all plaquettes of a 4D simple hypercubic lattice,
 $i_1,\,j_1,\,i_2$ and $j_2$ label the sites circulating about the 
plaquette and $U_{ji}$ is a $U(1)$ or a $SU(2)$ matrix ($N_c=1$ or 2) 
associated with the link $\langle ij\rangle$. The reversed link is 
associated with the inverse matrix. The aim is to calculate expectation 
values with respect to the Euclidean path integral 
\begin{equation} \label{SUN_Zk}
  Z = \int \prod_{\langle ij\rangle} dU_{ij}\,
      e^{+\beta_g\,S\left(\{U\}\right)}
\end{equation}
where the integrations are over the invariant group measure. While 
working at a particular link $\langle ij\rangle$, we need only to
consider the contribution to $S$, which comes from the staples 
containing this link. If we denote by $U_{\sqcup,k}$, $k=1,\dots,6$, 
the products which interact with the link in question, then the 
probability density of this link matrix is 
\begin{equation} \label{SUN_staples}
  dP(U) \sim dU\, \exp \left[ \frac{\beta_g}{N_c}\,{\rm Re}\, 
  {\rm Tr} \left( U \sum_{k=1}^6 U_{\sqcup,k} \right)\right]\ .
\end{equation}

\subsection{$SU(2)$}

We deal first with $SU(2)$ and parametrize the matrix elements in 
the form
\begin{equation} \label{SU2_parameters}
  U = a_0\,I + i\,\vec{a}\cdot \vec{\sigma},~~
  a_0^{\,2} + \vec{a}^{\,2} = 1,
\end{equation}
where $I$ denotes the $2\times 2$ identity matrix and $\vec{\sigma}$
are the Pauli matrices. A property of $SU(2)$ group elements is 
that any sum of them is proportional to another $SU(2)$ element. We 
define a $SU(2)$ matrix $U_{\sqcup}$ which corresponds to the sum of 
the staples in equation (\ref{SUN_staples}) by
\begin{equation} \label{SU2_staple}
  s_{\sqcup}\,U_{\sqcup} = \sum_{k=1}^6 U_{\sqcup,k},~~
  s_{\sqcup} = \sqrt{\det\left(\sum_{k=1}^6 U_{\sqcup,k}\right)}\ .
\end{equation}
Using the invariance of the group measure, one finds
\begin{equation} \label{SU2_hbmeasure}
  dP\left(U\,U^{-1}_{\sqcup}\right) \sim d\Omega\,da_0\,
  \sqrt{1-a_0^{~2}}\,\exp\left(\beta_g\,s_{\sqcup}\,a_0\right)
\end{equation}
where $d\Omega$ is the differential solid angle of $\vec{a}$. As it
is straightforward to generate the solid angle stochastically, the 
problem is reduced to sampling the probability density
\begin{equation} \label{Pa0}
  P(a_0) \sim \sqrt{1-a_0^{~2}}\,\exp\left(\beta_g\,s_{\sqcup}\,a_0\right)
\end{equation}
in the interval $-1\le a_0\le 1$. This is the starting point for 
the HBA, which amounts to finding a numerically fast inversion of 
the cumulative distribution function 
\begin{equation} \label{Fa0}
  F(a_0) = N_0 \int_{-1}^{a_0} da'_0\, \sqrt{1-a_0^{'~2}}\,
           \exp\left(\beta_g\,s_{\sqcup}\,a'_0\right)
\end{equation}
where $N_0$ ensures the normalization $F(1)=1$. The HBA updates 
$a_0$ by converting a uniformly distributed random number $0\le x<1$ 
into $a_0=F^{-1}(x)$.

The remark of our paper is that a crude tabulation of the function
$F(a_0)$ is entirely sufficient to obtain practically the same 
efficiency as with the HBA. Obviously, such a tabulation can still 
be done when there is no numerically efficient way to calculate 
$F^{-1}(x)$. The procedure does still generate the canonical 
probabilities of the continuous theory (\ref{SUN_Zk}) without any 
approximation (except by the floating point precision and limitations 
of the random number generator).

\begin{figure}[-t] \begin{center} % See subfolder su2_u1_rm2a/fig..
\epsfig{figure=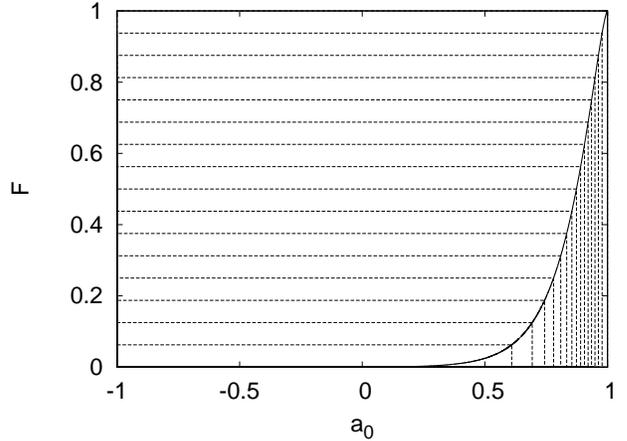,width=\columnwidth} \vspace{-1mm}
\caption{ Discretization of the cumulative distribution function 
$F(a_0;s_{\sqcup}^{11})$ for $SU(2)$ at $\beta_g=2.3$ for the choices 
$m=16$ (equidistant $s_{\sqcup}^i$ values, i.e., $s^{11}_{\sqcup}=
3.9375$) and $n=2^4=16$.  \label{fig_Fa0} }
\end{center} \vspace{-3mm} \end{figure}

Let us show how this works. First we choose a discretization of 
the parameter $s_{\sqcup}$, $0\le s_{\sqcup}\le 6$, into 
$m$ discrete values $s_{\sqcup}^i,\,i=1,\dots,m$ so that
\begin{equation} \label{s_order}
  0<s_{\sqcup}^1 < s_{\sqcup}^2 < \dots < s_{\sqcup}^m 
\end{equation}
holds. We take these values equidistant. Other partitions work too 
and could be more efficient. For each $s_{\sqcup}^i$ we calculate 
a table of values $a_0^{i,j},\, j=1,\dots, n$ defined by
\begin{equation} \label{a0_tab}
  \frac{j}{n} = F(a_0^{i,j};s_{\sqcup}^i)
\end{equation}
and we also tabulate the differences
\begin{equation} \label{dela0_tab}
  \triangle a_0^{i,j} = a_0^{i,j}-a_0^{i,j-1} 
  ~~{\rm for}~~j=1,\dots,n 
\end{equation}
where we define $a_0^{i,0}=-1$, and $a_0^{i,n}=+1$ follows from 
Eq.~(\ref{a0_tab}). For $\beta_g = 2.3$ this construction is shown 
in Fig.~\ref{fig_Fa0} using a representative $s_{\sqcup}^i$ value.

The biased Metropolis procedure for one update of a $SU(2)$ matrix 
consists now of the following steps:
% is now defined as follows:

\begin{enumerate}

\item Find the $s_{\sqcup}^i$ value (only $i$ is needed) which is 
      nearest to the actual $s_{\sqcup}$ value.

\item Place the present $a_0$ value on the discretization grid, 
      i.e., find the integer $j$ through the relation 
      $a_0^{i,j-1}\le a_0 < a_0^{i,j}$.

\item Pick an integer value $j'$ uniformly distributed in the range 
      $1$ to $n$.

\item Propose $a'_0=a_0^{i,j'-1}+x^r\,\triangle a_0^{i,j'}$, where 
      $x^r,\ 0\le x^r<1,$ is a uniformly distributed random number.

\item Accept $a'_0$ with the probability 
\begin{equation} \label{pacpt}
   p_a = { \exp (\beta_g s_{\sqcup} a'_0)\, \triangle a_0^{i,j'} 
   \over   \exp (\beta_g s_{\sqcup} a_0)\, \triangle a_0^{i,j} }\ .
\end{equation}

\item If $a'_0$ is accepted, calculate a random value for $\vec{a'}$ 
      with the measure $d\Omega$ and store the new $SU(2)$ matrix.
      Otherwise keep the old $SU(2)$ matrix. After this step the
      configuration has to be counted independently of whether $a'_0$ 
      was accepted or rejected.

\end{enumerate}

For $i$ given each interval on the $a_0$ abscissa of 
Fig.~\ref{fig_Fa0} is proposed with probability $1/n$. In the 
limit $n>m,\,m\to\infty$ these are by construction the heat bath 
probabilities, so that the acceptance rate becomes one. For a
reasonably accurate discretization the algorithm is still 
exact due to the factor $\triangle a_0^{i,j'}/\triangle a_0^{i,j}$ 
in the acceptance probability~(\ref{pacpt}), and the acceptance 
rate remains close to one. Therefore, the relative efficiency of 
a HBA versus our BMA becomes to a large extent a matter of CPU time 
consumption. 
% For the HBA the central point is whether an efficient numerical 
% inversion of the cumulative distribution function exists or not. 
% In contrast to that there is never a problem in using a 
% tabulation of this function.

\begin{figure}[-t] \begin{center} % See subfolder su2_u1_rm2a/fig..
\epsfig{figure=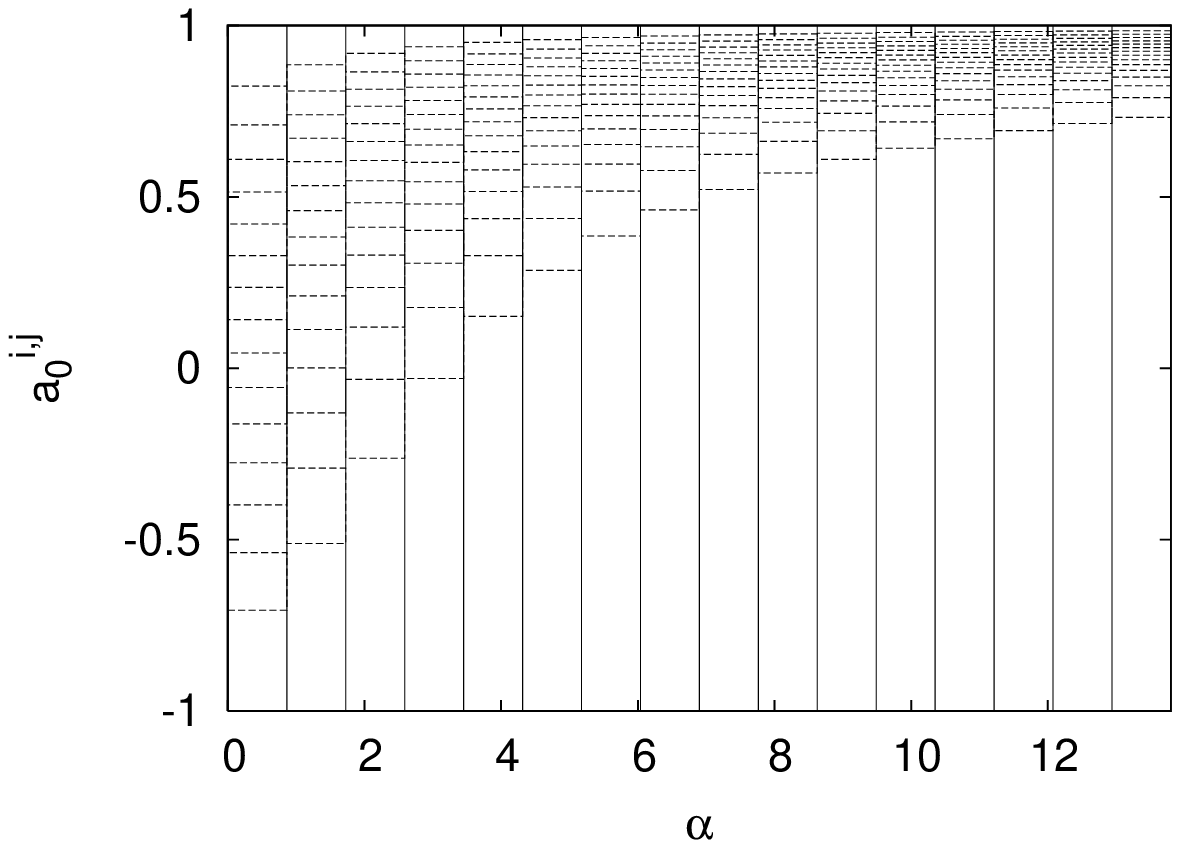,width=\columnwidth} \vspace{-1mm}
\caption{ Partition of the $\triangle a_0^{i,j}$ values for $SU(2)$ at 
$\beta_g=2.3$, where the variable $\alpha = \beta_g\,s_{\sqcup}$ is 
used on the abscissa. The choices for $m$ and $n$ are the same as 
in Fig.~\ref{fig_Fa0}.  \label{fig_su2tab} }
\end{center} \vspace{-3mm} \end{figure}

Only step~2 of the BMA procedure requires some thought, all others
are straightforward numerical calculations. For $n=2^{n_2}$ the 
interval label $j$ of the existing $a_0$ can be determined in $n_2$ 
steps using the binary search recursion
\begin{equation}  \label{recursion}
  j~\to~j\,+\,2^{i_2}\ {\rm sign}\,(\,a_0-a_0^{i,j}\,)\,,~~~i_2\to i_2-1 
\end{equation}
where the starting values are $i_2=n_2-2$ and $j=2^{n_2-1}$,
and the termination is for $i_2=0$ (after which one final logical
decision has to be made). As long as a uniform discretization of 
$s_{\sqcup}$ is chosen,  there is no slowing down of the code with 
an increase of the size $m$ of the table, while there is a logarithmic
slowing down with an increase of the $\triangle a_0^{i,j}$ 
discretization.  For the same choice of $m$ and $n$ values as 
used in Fig.~\ref{fig_Fa0}, the partition of all $\triangle a_0^{i,j}$ 
values is shown in Fig.~\ref{fig_su2tab}. For each bin $i$ on the 
abscissa the $a_0^{i,j}$ values are calculated for its central value 
$\alpha^i=\beta_g\,s^i_{\sqcup}$. For our simulations we used a 
finer discretization, $m=32$ and $n=128$.
% , which is unsuitable for illustration in a figure.

Table~\ref{tab_su2} illustrates the performance of the $SU(2)$ BMA 
for a long run on a $4\times 16^3$ lattice at $\beta_g=2.3$. At this 
coupling the system exhibits critical slowing down, because of its 
neighborhood to the deconfining phase transition (see for 
instance~\cite{FHK93} and references therein). Our comparison is with 
the Fabricius-Haan-Kennedy-Pendleton HBA \cite{FH84,KP85}, which at 
this coupling is more efficient than Creutz's HBA~\cite{Cr80}.

We used 16,384 sweeps for reaching equilibrium and, subsequently, 
$32\times 20,480$ sweeps for measurements. Simulations were performed 
on 2$\,$GHz Athlon PCs with the -O2 option of the (freely available) 
g77 Fortran compiler. Although our  programs are not thoroughly 
optimized, we report the runtimes in table~\ref{tab_su2}, because 
we expect their ratios to be relatively stable under further 
optimization. (For instance, our runs were fully in real*8 
precision. By reducing most of the code to real*4 a factor up 
to two might be gained.)

\begin{table}[tb]
\caption{ Efficiency of the $SU(2)$ algorithms on a $4\times 16^3$ 
lattice at $\beta_g=2.3$. For the same lattice size integrated 
autocorrelation times are also given at $\beta_g=2.2$ and 
$\beta_g=2.4$.  \label{tab_su2}}
% Results in sub-folder su2_u1_rm2a.
\medskip
\centering
\begin{tabular}{|c|c|c|} 
                     &HBA \cite{FH84,KP85}  & BMA          \\ \hline
CPU time             & 194,873 [s]          & 199,244 [s]   \\ \hline
Acceptance rate      & 1 (1.043 proposals)  & 0.975        \\ \hline
$\langle{\rm Tr}(U_{\Box})/2\rangle$ 
                     & 0.603147 (17)        &0.603111 (21) \\ \hline
$\tau_{\rm int}$     & 49.8 (3.5)           & 48.2 (3.8) \\ \hline
$\tau_{\rm int}(\beta=2.2)$& 7.1 (0.3)      &  8.9 (0.4) \\ \hline
$\tau_{\rm int}(\beta=2.4)$& 6.7 (0.4)      &  7.0 (1.0) 
\end{tabular} \end{table} \vspace*{0.2cm}

It is the high acceptance rate of 97.5\% which makes the BMA almost as 
efficient as the HBA. In standard Metropolis procedures one gets high 
acceptance rates only at the price of small moves, so that acceptance 
rates between 30\% and 50\% are optimal~\cite{bbook}. In our BMA the
high acceptance rate is achieved by proposing with an approximation of
heat bath probabilities for which the acceptance rate is 100\%. So, an 
acceptance rate close to 100\% is best for the BMA. The accept/reject 
step corrects for the failure to approximate the heat bath probability 
perfectly. 

Although the acceptance rate for the HBA is 100\%, the $SU(2)$ HBAs 
use in their inner loops a reject until accepted (RUA) step. It should 
be noted that this is distinct from the accept/reject step of the 
BMA. Like in the original Metropolis method, the latter cannot be 
iterated until accepted (compare, e.g., p.137 of Ref.~\cite{bbook}). 
This would introduce an uncontrolled bias, which for the original 
Metropolis algorithm is towards too low energies. For the simulation 
of table~\ref{tab_su2} the RUA step of the HBA \cite{FH84,KP85} needs 
in the average 1.043 iterations to generate the new $a_0$ 
value~\cite{acpt}. For small $\beta_g$ values the number of iterations
goes up, so that the Creutz HBA becomes then more efficient than the
HBA of Fabricius-Haan-Kennedy-Pendleton, see \cite{KP85} for a detailed 
discussion. Independently of $\beta_g$ the BMA acceptance rate stays 
always close to 100\%. 

The difference between a RUA procedure and the accept/reject step of 
a BMA becomes important for a (checkerboard) parallelization of the 
updating. While for a BMA the speed is uniform at all nodes, this 
is not the case for a RUA method, where all nodes have to wait until 
the last RUA step is completed. For large systems, the consequences 
would be disastrous, so that at the price of an arguably negligible 
bias workers tend to impose an upper limit on the number of RUA steps 
(say three for our $SU(2)$ case).

The integrated autocorrelation time $\tau_{\rm int}$ is a direct 
measure for the performance of an algorithm. The number of sweeps 
needed to achieve a desired accuracy is directly proportional to
$\tau_{\rm int}$. Table~\ref{tab_su2} gives $\tau_{\rm int}$ for 
the Wilson plaquette together with the expectation value of this 
operator. Error bars are given in parenthesis and apply to the last 
digits. They are calculated with respect to 32 bins (jackknife bins 
in case of $\tau_{\rm int}$), relying on the data analysis software 
of~\cite{bbook}. We see that the expectation values are well 
compatible with one another ($Q=0.18$ in a Gaussian difference 
test). For $\tau_{\rm int}$ we know that the HBA should give a 
slightly lower value than the BMA. That the $\tau_{\rm int}$ data 
at $\beta_g=2.3$ table come out in the opposite order is attributed 
to a statistical fluctuation. This is confirmed by shorter runs which 
we performed at other $\beta_g$ values, whose $\tau_{\rm int}$ results 
are also listed in the table.

\subsection{$U(1)$}

Next we consider the $U(1)$ gauge group. The ``matrices'' are then 
complex numbers on the unit circle, $U_{ij}=\exp (i\,\phi_{ij})$, and 
the analogue of Eq.~(\ref{SU2_staple}) becomes
\begin{equation}  \label{U1_staple}
  r_{\sqcup}\,e^{i\phi_{\sqcup}} = \sum_{k=1}^6 e^{i\phi_{\sqcup,k}}\,,
\end{equation}
$ r_{\sqcup} = \sqrt{ \left(\sum_{k=1}^6 \cos\phi_{\sqcup,k}\right)^2
   + \left(\sum_{k=1}^6 \sin\phi_{\sqcup,k}\right)^2  } $.
We are led to the cumulative distribution function
\begin{equation}  \label{Fphi}
  F_1(\phi) = N_1 \int_0^{\phi} d\phi'\,e^{\beta_g\,r_{\sqcup}\,
  \cos (\phi')} 
\end{equation}
where the normalization is $F_1(2\pi)=1$ and the angle 
$(\phi+\phi_{\sqcup})\,{\rm mod}(2\pi)$ will be stored. 

We test the performance of the $U(1)$ BMA for a $4\times 16^3$ lattice 
at $\beta_g=1.0$, again a coupling which puts the system close to the 
deconfining phase transition, which is weakly first order for $U(1)$ 
(see for instance~\cite{ASL99} and references therein). HBAs have been 
designed in Ref.~\cite{We89,HaNa92}. Both HBAs rely on a RUA step, so 
that the remarks made in this connection for $SU(2)$ apply
also to $U(1)$. We have only tested the HBA of Ref.~\cite{HaNa92}, 
which turns out to be about 20\% slower than our BMA, while the 
integrated autocorrelation time is about 10\% lower. Overall an 
advantage of 10\% in favor of the $U(1)$ BMA, which re-iterates that 
HBAs and BMAs have about equal efficiency, when efficient HBAs exist.

\begin{figure}[-t] \begin{center}
\epsfig{figure=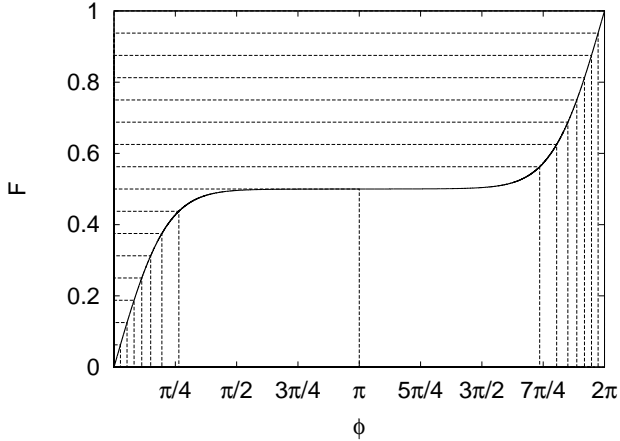,width=\columnwidth} \vspace{-1mm}
\caption{ Discretization of the cumulative distribution function 
$F(\phi;r_{\sqcup}^{11})$ for $U(1)$ at $\beta_g=1.0$ for the choices 
$m=16$ (equidistant $r_{\sqcup}^i$ values) and $n=2^4=16$.  
\label{fig_Fphi} }
\end{center} \vspace{-3mm} \end{figure}

\begin{figure}[-t] \begin{center}
\epsfig{figure=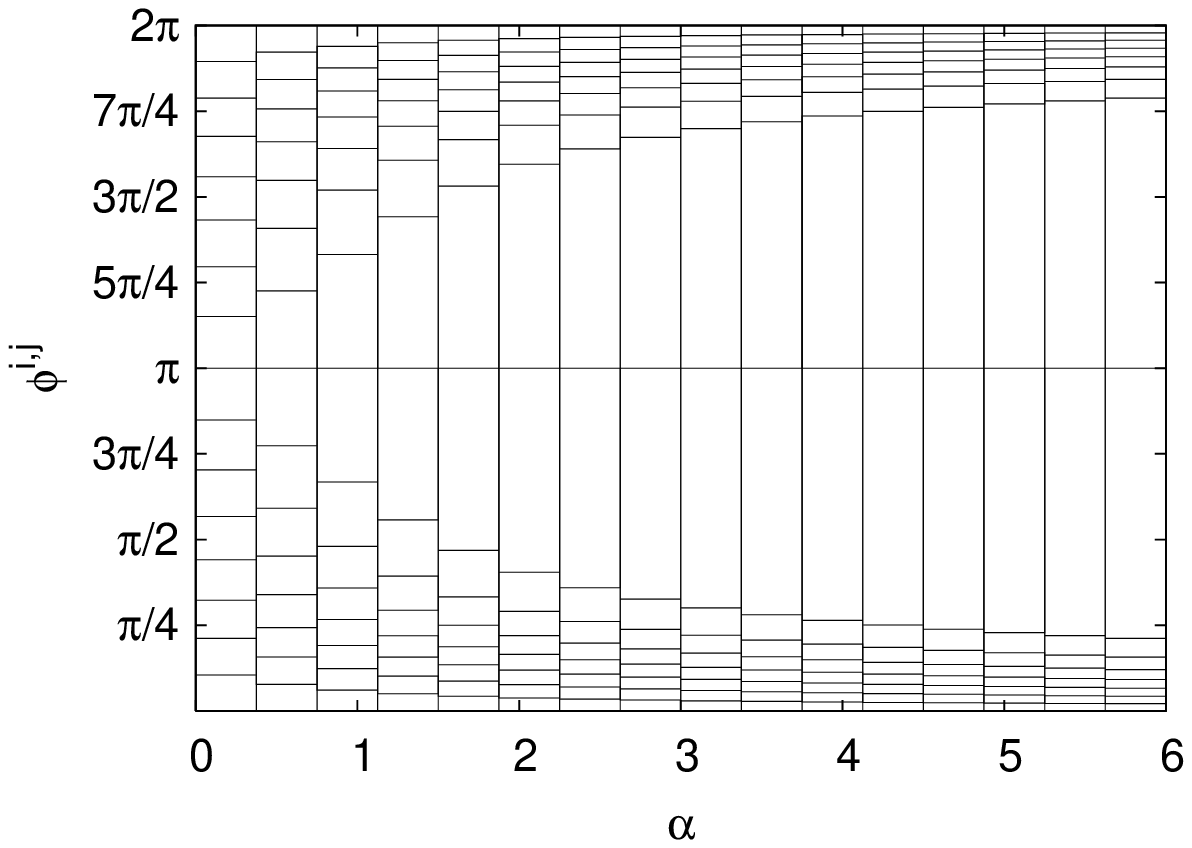,width=\columnwidth} \vspace{-1mm}
\caption{ Partition of the $\triangle \phi^{i,j}$ values for $U(1)$ at 
$\beta_g=1.0$, where the variable $\alpha=\beta_g\,r_{\sqcup}$ is used 
on the abscissa. The choices for $m$ and $n$ are the same a in 
Fig.~\ref{fig_Fphi}.  \label{fig_u1tab} }
\end{center} \vspace{-3mm} \end{figure}

We compare the $U(1)$ BMA now with a conventional Metropolis 
algorithm, which proposes new angles uniformly in the (entire) range 
$[0,2\pi)$. For the BMA we follow the same lines as previously for $
F(a_0)$ of Eq.~(\ref{Fa0}). Fig.~\ref{fig_Fphi} plots $F_1(\phi)$ at 
$\beta_g=1.0$ using a representative $r^i_{\sqcup}$ value and 
Fig.~\ref{fig_u1tab} shows the entire tabulation $\triangle\phi^{i,j}$.
Table~\ref{tab_u1} summarizes the results. At $\beta_g=1$ the 
acceptance rate of the standard Metropolis procedure is still 
about 30\%, so that a restriction of the proposal range to 
increase the acceptance rate is not warranted~\cite{bbook}. 
From the data of the table we conclude that the BMA improves 
the Metropolis performance at $\beta_g=1$ by a factor of about two. 

When comparing with a full-range Metropolis algorithm an upper bound 
on the improvement factor is given by the ratio of the acceptance 
rates, in the present case $0.972/0.282 = 3.45$. This applies also 
to comparisons of such Metropolis algorithms with HBAs, substituting 
then one for the acceptance rate. The bound will normally not be 
saturated, because rms deviations of the new variables from the old 
variables are smaller for a BMA or HBA than for a full-range Metropolis 
algorithm. Larger gains can be achieved when the Metropolis acceptance 
rates are small. For $U(1)$ this happens for $\beta_g\gg 1$. 

\begin{table}[tb]
\caption{ Efficiency of the $U(1)$ algorithms on a $4\times 16^3$ 
lattice at $\beta_g=1.0$. \label{tab_u1}}
% Results in sub-folder su2_u1_rm2a.
\medskip
\centering
\begin{tabular}{|c|c|c|} 
                     & Metropolis     & BMA          \\ \hline
CPU time             &  84,951 [s]    & 107,985 [s]  \\ \hline
Acceptance rate      &  0.286         & 0.972        \\ \hline
$\langle\cos\phi_{\Box}\rangle$ 
                     & 0.59103 (16)   &0.59106 (12)  \\ \hline
$\tau_{\rm int}$     & 341 (26)       & 142 (10)     \\
\end{tabular} \end{table} \vspace*{0.2cm}

\section{Summary and Conclusions}

In summary, BMAs are an alternative to HBAs. BMAs work still in 
situations for which HBAs fail, because there is no efficient inversion 
of the cumulative distribution function in question. In lattice gauge 
theory this is the case for some Higgs system and for actions which 
are non-linear in the Wilson plaquette operator (see, e.g., 
Ref.~\cite{DaHe88} and references therein). Obviously, similar 
situations ought to exist for energy functions in many other fields. 
We leave it to the reader to identify whether her or his simulations 
would benefit from using a BMA.
Finally, let us mention that BMAs may be combined with overrelaxation 
moves \cite{Ad81,BrWo87,Cr87} in the same way as one does for HBAs or 
standard Metropolis algorithms.
% One may want to denote the BMA discussed in this paper by
% BM1.1 (Biased Metropolis 1.1), where in a BM$n.m$ notation $n$ 
% counts the number of update variables used and $m$ the number of
% parameters. Following lines similar to those discussed in 
% Ref.~\cite{BZ05} it appers to be feasible to extent BMAs up to
% using two variables and two parameters. This is not possible 
% for the HBAs, because the numerical inversion of cumulative 
% distribution function of several variables is certainly impractical.
% The main difference between the BMAs discussed in this paper and
% the Rugged Metropolis (RM) approach of Ref.~\cite{Be03,BZ05} is
% that the RM is a particular scheme, which uses numerical information 
% from higher temperatures to feed it into a BM simulation at a lower 
% temperature. In contrast to that all BM probabilities used in
% this paper are calculated from first principles without using
% any input from previous simulations.  

\acknowledgments
This work was in part supported by the US Department of Energy 
under contract DE-FG02-97ER41022. We like to thank Urs Heller 
for useful correspondence.

\clearpage

\begin{thebibliography}{19}
\bibitem{acpt} In some of the literature the quantity 1/(average number
of RUA heat bath iterations per update) is is also called acceptance 
rate. It should not be confused with the acceptance rate defined by 
our Eq.~(\ref{pacpt}).
\end{thebibliography}
\end{document}